\documentclass[twocolumn,showpacs,showkeys,preprintnumbers]{revtex4-1}
\usepackage{amssymb}

\usepackage{amsfonts}

\usepackage{latexsym}

\usepackage{dcolumn}

\usepackage{amsmath}

\usepackage{graphicx}

\usepackage{psfrag,pstricks}

\begin{document}

\title{Intrinsic cross-Kerr nonlinearity in an optical cavity containing an interacting Bose-Einstein condensate}

\author{A. Dalafi$^{1}$ }
\email{adalafi@yahoo.co.uk}

\author{M. H. Naderi$^{2}$}

\affiliation{$^{1}$ Laser and Plasma Research Institute, Shahid Beheshti University, Tehran 19839-69411, Iran\\
$^{2}$Quantum Optics Group, Department of Physics, Faculty of Science, University of Isfahan, Hezar Jerib, 81746-73441, Isfahan, Iran}

\date{\today}

\begin{abstract}
An interacting cigar-shaped Bose-Einstein condensate (BEC) inside a driven optical cavity exhibits an intrinsic cross-Kerr (CK) nonlinearity due to the interaction with the optical mode of the cavity. Although the CK coupling is much weaker than those of the radiation pressure and the atom-atom interactions, it can affect the bistability behavior of the system when the intensity of the laser pump is strong enough. On the other hand, there is a competition between the CK nonlinearity and the atom-atom interaction so that the latter can neutralize the effect of the former. Furthermore, the CK nonlinearity causes the effective frequency of the Bogoliubov mode of the BEC as well as the quantum fluctuations of the system to be increased by increasing the cavity driving rate. However, in the dispersive interaction regime the effect of the CK nonlinearity is negligible. In addition, we show that by increasing the \textit{s}-wave scattering frequency of atomic collisions one can generate a strong stationary quadrature squeezing in the Bogoliubov mode of the BEC.

\end{abstract}


\maketitle
\section{Introduction}
%
%
Hybrid systems consisting of Bose-Einstein condensates (BEC) inside optical cavities exhibit by themselves optomechanical proprties \cite{Brenn Science,Ritter Appl. Phys. B,Gupta,Brenn Nature}.  The excitation of a collective mode of BEC couples to the radiation pressure of the cavity optical field just like what happens for the moving mirror in a bare optomechanical system \cite{Kanamoto 2010,Szimai2010,dalafi1}. One of the most important features of such systems is their nonlinear proprties like the Kerr effect \cite{Nori2009,Meystre85,Mancini94}.

Recently, it has been proposed \cite{Heikkila} and experimentally demonstrated \cite{Pirkkalainen} that a cross-Kerr (CK) type of coupling between the moving mirror and the optical field can be realized in a bare optomechanical cavity which is coupled to a Josephson junction. The CK nonlinearity leads to a frequency shift in both the mechanical and optical modes\cite{Khan} and also affects the stability of the optomechanical system \cite{Sarala}. In addition, it can change the optical bistability of the optomechanical system into a tristable behavior \cite{Xiong}.

On the other hand, a hybrid system consisting of a BEC traped inside an optical cavity exhibits an intrinsic CK nonlinearity if the cavity is driven by a laser pump which is far detuned from the atomic resonance. In spite of bare optomechanical systems, in such hybrid systems there is no need to use any aditional component (like the Josephson junction) to generate the CK coupling because here the CK nonlinearity emerges naturally due to the direct interaction between the atoms and the optical field of the cavity.
 
In fact, there exist two kinds of coupling between the Bogoliubov mode of the BEC and the optical field: the radiation pressure and the CK coupling. However, the ratio of the CK to the optomechanical coupling is of the order of $ \frac{1}{\sqrt{N}} $ where $ N $ is the number of atoms. Therefore, in many cases \cite{Singh,nagy2013,Chiara,Rogers,Zhang 2009} one can neglect the CK term in comparison to the radiation pressure interaction. That is why we have also not considered the effect of this term on the dynamics of the system in our previous papers \cite{dalafi2,dalafi3,dalafi5,dalafi6}. In addition to the radiation pressure and CK nonlinearities, there is another kind of nonlinearity which is due to the atom-atom interaction in the BEC which may lead to some interesting phenomena such as the quantum phase transition \cite{dalafi4}.

In this work, we are going to investigate under which conditions and in which regimes the effect of CK nonlinearity manifest itself in a hybrid system consisting of a BEC traped inside an optical cavity. We show that it can change the optical bistability behavior of the system specially when the driving rate of the optical cavity is increased. It is also shown that the effective frequency of the Bogoliubov mode of the BEC depends on the mean value of the optical mode through the CK coupling so that the Bogoliubov mode frequency  gets larger by increasing the number of intracavity photons. This is one of the main results of our investigation.

Moreover, we show that the CK effects get very small in the dispersive interaction regime where the effective cavity detunig is much larger than the damping rate of the cavity.  However, it increases the quantum fluctuations near the bistability region and can also strengthen the atom-field entanglement in a wide range of the cavity detuning. On the other hand, we investigate the nonlinear effect of atom-atom interaction against the CK nonlinearity and show how it neutralizes the CK effects. More importantly, it is shown that in the dispersive regime where the CK effect disappears, a steady-state quadrature squeezing in the Bogoliubov mode of the BEC can be generated by increasing the \textit{s}-wave scattering frequency of the atom-atom interaction.

The structure of the paper is as follows. In Sec. II we derive the Hamiltonian of the system consisting of a BEC inside an optical cavity, and in Sec. III the dynamics of the system is described. In Sec. IV we study the CK effect on the optical bistability, and in Sec.V we investigate the CK effect on the entanglement and squeezing. Finally, our conclusions are summarized in Sec. VI.

\section{System Hamiltonian}
\begin{figure}[ht]
\centering
\includegraphics[width=3in]{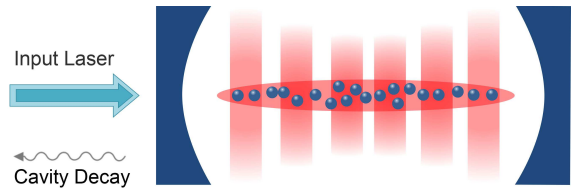} 
\caption{(Color online) A BEC trapped inside an optical cavity interacting with a single cavity mode. The cavity mode is driven by a laser at rate $\eta$ and the cavity decay rate is $\kappa$.}
\label{fig:fig1}
\end{figure}

As depicted in Fig.\ref{fig:fig1}, the system we are going to investigate is an optical cavity with length $L$ consisting of a cigar-shaped BEC of $N$ two-level atoms with mass $m_{a}$ and transition frequency $\omega_{a}$ confined in a cylindrically symmetric trap with a transverse trapping frequency $\omega_{\mathrm{\perp}}$ and negligible longitudinal confinement along the $x$ direction. An external laser with frequency $\omega_{p}$ and wave number $k=\omega_{p}/c$  drives the cavity at rate $\eta=\sqrt{2\mathcal{P}\kappa/\hbar\omega_{p}}$ through one of its mirrors ($\mathcal{P}$ is the laser power and $\kappa$ is the cavity decay rate).

If the laser pump is far detuned from the atomic resonance ($\Delta_{a}=\omega_{p}-\omega_{a}$  exceeds the atomic linewidth $\gamma$ by orders of magnitude), the excited electronic state of the atoms can be adiabatically eliminated and the atomic spontaneous emission can be neglected \cite{Masch Ritch 2004,Dom JB,Nagy Ritsch 2009}. In this way, the dynamics of the system can be described within an effective one-dimensional model by quantizing the atomic motional degree of freedom along just the $x$ axis. In the frame rotating at the pump frequency, the many-body Hamiltonian of the system reads
\begin{eqnarray}\label{H1}
H&=&\int_{-L/2}^{L/2} dx \Psi^{\dagger}(x)\Big[\frac{-\hbar^{2}}{2m_{a}}\frac{d^{2}}{dx^{2}}+\hbar U_{0} \cos^2(kx) a^{\dagger} a\nonumber\\
&&+\frac{1}{2} U_{s}\Psi^{\dagger}(x)\Psi(x)\Big] \Psi(x)-\hbar\Delta_{c} a^{\dagger} a + i\hbar\eta(a - a^{\dagger}),\nonumber\\
\end{eqnarray}
where $ \Psi(x) $ and $ a $ are, respectively, the annihilation operators of the atomic field and the cavity mode. In addition, $ \Delta_{c}=\omega_{p}-\omega_{0} $ is the cavity-pump detuning where $ \omega_{0} $ is the resonance frequency of the cavity. $U_{0}=g_{0}^{2}/\Delta_{a}$ is the optical lattice barrier height per photon which represents the atomic backaction on the field, $g_{0}$ is the vacuum Rabi frequency, $U_{s}=\frac{4\pi\hbar^{2} a_{s}}{m_{a}}$ and $a_{s}$ is the two-body \textit{s}-wave scattering length \cite{Masch Ritch 2004,Dom JB}. 

If the intracavity photon number is so low that the condition $U_{0}\langle a^{\dagger}a\rangle\leq 10\omega_{R}$ is satisfied where $\omega_{R}=\frac{\hslash k^{2}}{2m_{a}}$ is the recoil frequency of the condensate atoms, and under the Bogoliubov approximation \cite{Nagy Ritsch 2009}, the atomic field operator can be expanded as the following single-mode quantum field
\begin{equation}\label{opaf}
\Psi(x)=\sqrt{\frac{N}{L}}+\sqrt{\frac{2}{L}}\cos(2kx) c,
\end{equation}
where the first term is the condensate mode which is considered as a c-number and the operator $ c $ in the second term (the so-called Bogoliubov mode) corresponds to the quantum fluctuations of the atomic field about the classical condensate mode. By substituting the atomic field operator of Eq.(\ref{opaf}) into Eq.(\ref{H1}), we can find the Hamiltonian of the system in the following form
\begin{eqnarray}\label{Hc}
H&=&\hbar\delta_{c} a^{\dagger} a + i\hbar\eta(a-a^{\dagger})+\hbar\Omega_{c} c^{\dagger}c+\hbar\zeta a^{\dagger}a (c+c^{\dagger})\label{subH}\nonumber\\
&&+\frac{1}{4}\hbar\omega_{sw}(c^{2}+c^{\dagger 2})+\hbar g a^{\dagger}a c^{\dagger}c.
\end{eqnarray}
Here, $ \delta_{c}=-\Delta_{c}+\frac{1}{2}N U_{0} $ is the Stark-shifted cavity frequency due to the presence of the BEC, $ \Omega_{c}=4\omega_{R}+\omega_{sw} $ is the frequency of the Bogoliubov mode, $ \zeta=\frac{\sqrt{2N}}{4}U_{0} $ is the optomechanical coupling between the Bogoliubov and the optical modes, $ \omega_{sw}=8\pi\hbar a_{s}N/m_{a}Lw^2 $ is the \textit{s}-wave scattering frequency of the atomic collisions ($ w $ is the waist radius of the optical mode), and $ g=\frac{1}{2}U_{0} $ is the CK coupling.

As is seen from Eq.(\ref{subH}) there are three kinds of nonlinearity in the Hamiltonian of the system: the radiation pressure interaction between the Bogoliubov mode and the optical field (the fourth term), the nonlinear atom-atom interaction (the fifth term),  and the CK nonlinearity (the sixth term). The important point is that the manifestation of the nonlinear CK term in the Hamiltonian of the system is due to the direct atom-photon interaction. In other words, it is an intrinsic property of the system.

A comparison between the above-mentioned three nonlinear terms show that the radiation pressure interaction is the most important term in the system because of the large value of the optomechanical coupling parameter $ \zeta $. The strength of the atom-atom interaction is determined by the \textit{s}-wave scattering frequency $ \omega_{sw} $ whose value can be increased from zero up to several tens of $ \omega_{R} $ by manipulating the transverse trapping frequency $\omega_{\mathrm{\perp}}$ which can change the waist radius of the optical mode $ w $ \cite{Morsch}.The CK term whose strength is determined by the CK coupling parameter $ g=U_{0}/2 $ has the weakest effect among the others. In fact the ratio of the CK coupling to the optomechanical coupling is of the order of $ \frac{1}{\sqrt{N}} $. So, for very large values of $ N $ the CK term is negligible in comparison to the radiation pressure and atom-atom interactions. That is why in the previous works \cite{dalafi2,dalafi3,dalafi5,dalafi6} this term has been disregarded.

\section{Dynamics of the system}
The dynamics of the system described by the Hamiltonian in Eq.(\ref{subH}) is fully characterized by the following set of nonlinear Heisenberg-Langevin equations
\begin{eqnarray}\label{NHL}
\dot{a}&=&-(i\delta_{c}+\kappa)a-i\zeta a(c+c^{\dagger})-ig a c^{\dagger}c-\eta+\sqrt{2\kappa}\delta a_{in},\nonumber\\
\dot{c}&=&-(i\Omega_{c}+\gamma)c-\frac{i}{2}\omega_{sw}c^{\dagger}-i\zeta a^{\dagger}a-ig a^{\dagger}a c+\sqrt{2\gamma}\delta c_{in}.\nonumber\\
\end{eqnarray}
Here $ \gamma $ characterizes the dissipation of the Bogoliubov mode of the BEC. The optical field quantum vacuum fluctuation $\delta a_{in}(t)$ satisfies the Markovian correlation functions, i.e., $\langle\delta a_{in}(t)\delta a_{in}^{\dagger}(t^{\prime})\rangle=(n_{ph}+1)\delta(t-t^{\prime})$, $\langle\delta a_{in}^{\dagger}(t)\delta a_{in}(t^{\prime})\rangle=n_{ph}\delta(t-t^{\prime})$ with the average thermal photon number $n_{ph}$ which is nearly zero at optical frequencies \cite{Gardiner}. Furthermore, $\delta c_{in}(t)$ is the quantum noise input for the Bogoliubov mode of the BEC which also satisfies the same Markovian correlation functions as those of the optical noise \cite{K Zhang}. The noise sources are assumed uncorrelated for the different modes of both the matter and light fields.

In order to linearize the nonlinear set of equations (\ref{NHL}) one can expand the quantum operators  around their respective classical mean values as $ a=\alpha+\delta a $ and $ c=\beta+\delta c $ where $ \delta a $ and $ \delta c $ are small quantum fluctuations around the mean fields $ \alpha $ and $ \beta $. In this way, a set of nonlinear algebraic equations for the mean-field values and another set of linear ordinary differential equations for the quantum fluctuations will be obtained. The steady-state mean-field values are obtained as follows
\begin{subequations}
\begin{eqnarray}
\alpha&=&-\frac{\eta}{i\Delta+\kappa},\\
\beta&=&-\zeta|\alpha|^{2}\frac{\Omega^{(-)}+i\gamma}{\Omega^{(+)}\Omega^{(-)}+\gamma^{2}},
\end{eqnarray}
\end{subequations}
where $ \Delta=\delta_{c}+2\beta_{R}\zeta+g|\beta|^{2} $ with $ \beta_{R} $ being the real part of the complex mean field $ \beta $, and $ \Omega^{(\pm)}=\Omega_{c}\pm\frac{1}{2}\omega_{sw}+g|\alpha|^{2} $. 

On the other hand, by introducing the optical field and the Bogoliubov mode quadratures as
\begin{subequations}\label{quadratures}
\begin{eqnarray}
\delta X&=&\frac{1}{\sqrt{2}}(\delta a+\delta a^{\dagger}),  \delta Y=\frac{1}{\sqrt{2}i}(\delta a-\delta a^{\dagger}),\\
\delta Q&=&\frac{1}{\sqrt{2}}(\delta c+\delta c^{\dagger}),  \delta P=\frac{1}{\sqrt{2}i}(\delta c-\delta c^{\dagger}),
\end{eqnarray}
\end{subequations}
the following linearized set of ordinary differential equations is obtained for the quantum fluctuations
\begin{equation}\label{nA}
\delta\dot{u}(t)=A \delta u(t)+\delta n(t),
\end{equation}
where $\delta u(t)=[\delta X,\delta Y,\delta Q,\delta P]^{T}$ is the vector of continuous variable fluctuation operators and
$ \delta n(t)=[\sqrt{2\kappa}\delta X_{in},\sqrt{2\kappa}\delta Y_{in},\sqrt{2\gamma}\delta Q_{in},\sqrt{2\gamma}\delta P_{in}]^{T} $ is the corresponding vector of noises. The $4\times4$ matrix $A$ is the drift matrix given by
\begin{equation}\label{A}
A=\left(\begin{array}{cccc}
-\kappa & \Delta & G_{I} & F_{I} \\
   -\Delta & -\kappa &-G_{R} &-F_{R} \\
    F_{R} & F_{I} & -\gamma & \Omega^{(-)} \\
    -G_{R} & -G_{I} & -\Omega^{(+)} & -\gamma\\
    \end{array}\right),
\end{equation}
where the new parameters in the drift matrix of Eq.(\ref{A}) have been defined in terms of the real and imaginary parts of the optical and atomic mean fields as follows
\begin{subequations}
\begin{eqnarray}
G_{R}&=&2\alpha_{R}(\zeta+g\beta_{R}),\\
 G_{I}&=&2\alpha_{I}(\zeta+g\beta_{R}),\\
F_{R}&=&2g\alpha_{R}\beta_{I},\\
F_{I}&=&2g\alpha_{I}\beta_{I}.
\end{eqnarray}
\end{subequations}

The solutions to Eq.(\ref{nA}) are stable only if all the eigenvalues of the matrix $ A $ have negative real parts. The stability conditions can be obtained, for example, by using the Routh-Hurwitz criteria \cite{RH}.

Based on the dynamical equations (\ref{nA}), the quadratures of the Bogoliubov mode of the BEC oscillate effectively with the frequency $ \omega_{B}=\sqrt{\Omega^{(+)}\Omega^{(-)}} $ which is given by
\begin{equation}\label{wB}
\omega_{B}=\sqrt{\Big(4\omega_{R}+\frac{1}{2}\omega_{sw}+g|\alpha|^{2}\Big) \Big(4\omega_{R}+\frac{3}{2}\omega_{sw}+g|\alpha|^{2}\Big)}.
\end{equation}

Based on Eq.(\ref{wB}), the effective frequency of the Bogoliubov mode depends not only on the \textit{s}-wave scattering frequency of the atom-atom interaction ($ \omega_{sw} $) but also on the mean number of interacavity photons ($ |\alpha|^{2} $) which is due to the presence of the CK nonlinearity. In the absence of the CK nonlinearity ($ g=0 $), the effective frequency of the Bogoliubov quadrature reduces to the Bogoliubov frequency in the absence of the CK nonlinearity $ \omega_{c}=\omega_{B}|_{g=0} $ which had been obtained in the previous works \cite{dalafi2,dalafi3,dalafi5,dalafi6}.

\section{effect of the CK nonlinearity on optical bistability}

\begin{figure}[ht]
\centering
\includegraphics[width=2.8 in]{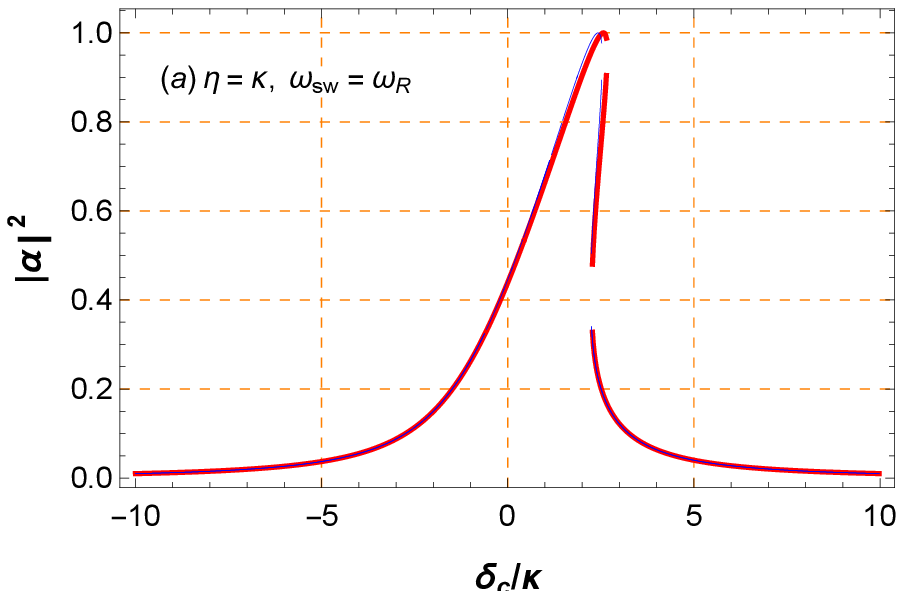}
\includegraphics[width=2.8 in]{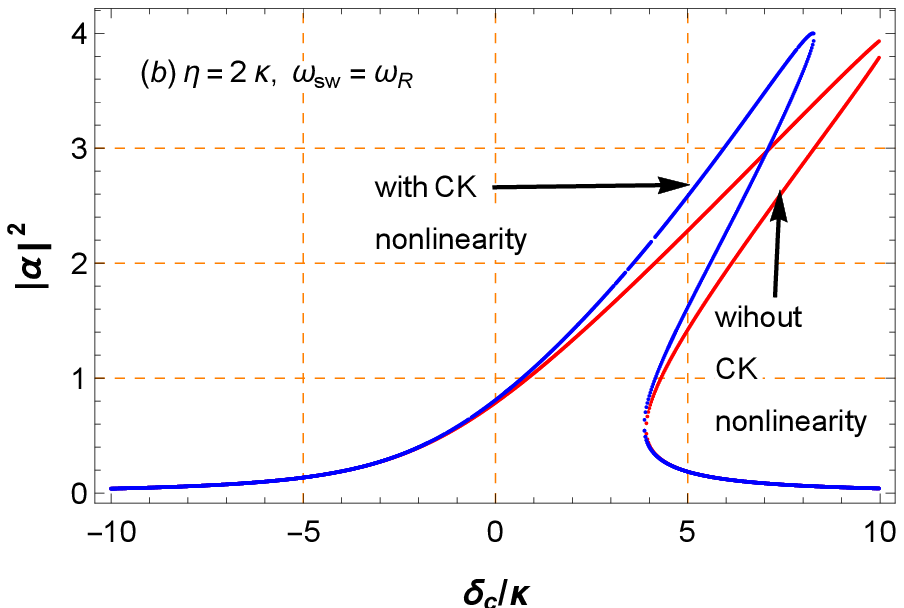}
\caption{
(Color online) The mean number of intracavity photons versus the normalized detuning $ \delta_{c}/\kappa $ for two values of the external laser driving rate: (a)  $ \eta=\kappa $ and (b)  $ \eta=2\kappa $ when $ \omega_{sw}=\omega_{R} $. The red (blue) line corresponds to the absence (presence) of the CK nonlinearity. The other parameters are  $ L=187 \mu$m, $ \lambda=780 $nm and $ \kappa=2\pi\times 1.3 $MHz. The cavity contains $ N=10^5 $ Rb atoms with $ \omega_{sw}=\omega_{R} $ and $ \gamma_{c}=0.001\kappa $.}
\label{fig:fig2}
\end{figure}

In this section we show how the CK nonlinearity affects the bistability behavior of the system. For this purpose we analyse our results based on the experimentally feasible parameters given in \cite{Ritter Appl. Phys. B, Brenn Science},i.e., we assume there are $ N=10^5 $ Rb atoms inside an optical cavity of length $ L=187 \mu$m with bare frequency $ \omega_{0}=2.41494\times 10^{15} $Hz corresponding to a wavelength of $ \lambda=780 $nm. The atomic $ D_{2} $ transition corresponding to the atomic transition frequency $ \omega_{a}=2.41419\times 10^{15} $Hz couples to the mentioned mode of the cavity. The atom-field coupling strength $ g_{0}=2\pi\times 14.1 $MHz and the recoil frequency of the atoms is $ \omega_{R}=23.7 $KHz. Furthermore, we assume that the equilibrium temperature of the BEC is $ T=0.1\mu $K. For these set of experimental data the the CK coupling parameter is three orders of magnitude smaller than the optomechanical coupling parameter $ (g\approx 0.004 \zeta) $.

In Fig.\ref{fig:fig2} we have plotted the mean number of intracavity photons versus the normalized detuning $ \delta_{c}/\kappa $ for two values of the external laser driving rate $ \eta=\kappa $ [Fig.\ref{fig:fig2}(a)] and $ \eta=2\kappa $ [Fig.\ref{fig:fig2}(b)] when the \textit{s}-wave scattering frequency has been fixed at $ \omega_{sw}=\omega_{R} $. The red (blue) line corresponds to the absence (presence) of the CK nonlinearity. As is seen from Fig.\ref{fig:fig2}(a) for a driving rate of $ \eta=\kappa $ the two curves (red and blue lines) overlap with each other completely. It means that for small driving rates the presence of the CK nonlinearity cannot affect the bistability of the system. However, by increasing the intensity of the pump laser at a fixed value of $ \omega_{sw} $ the effect of the CK nonlinearity manifest itself specifically in the bistabilty region. This can be seen in Fig.\ref{fig:fig2}(b) where the two curves are resolved from each other in the bistability region.

\begin{figure}[ht]
\centering
\includegraphics[width=2.8 in]{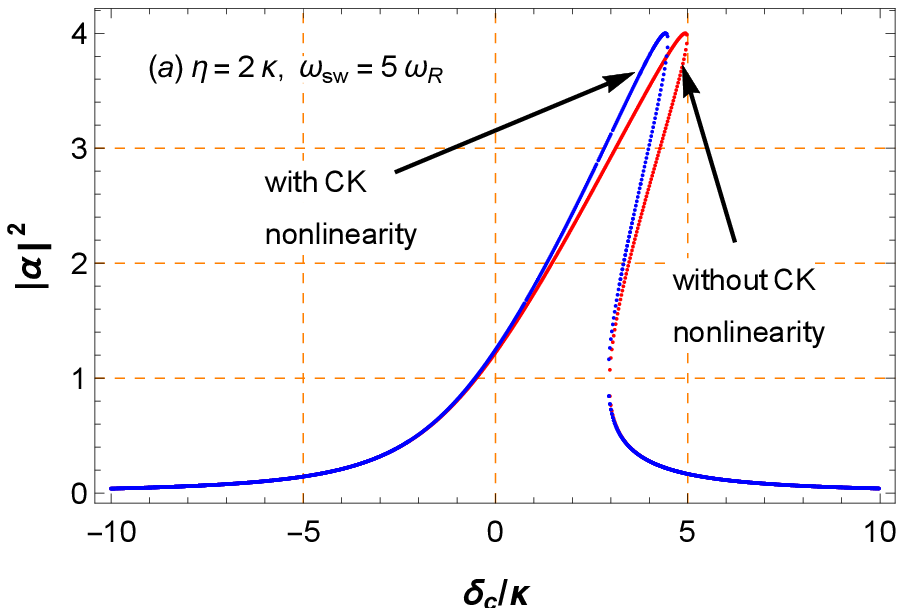}
\includegraphics[width=2.8 in]{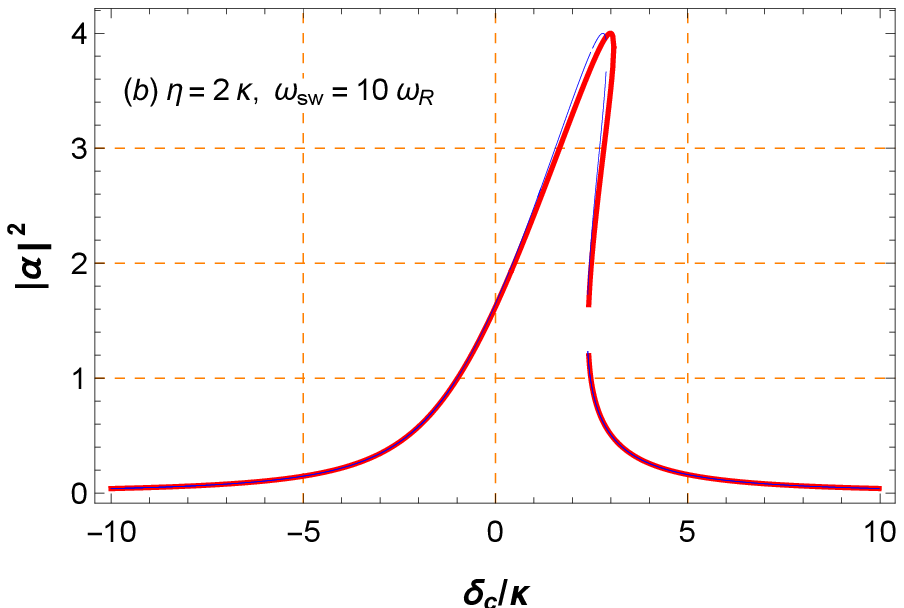}
\caption{
(Color online) The mean number of intracavity photons versus the normalized detuning $ \delta_{c}/\kappa $ for two values of the \textit{s}-wave scattering frequency: (a)  $\omega_{sw}=5\omega_{R} $ and (b)  $ \omega_{sw}=10\omega_{R} $ when the pump rate is $ \eta=2\kappa $. The red (blue) line corresponds to the absence (presence) of the CK nonlinearity. The other parameters are the same as those in Fig.\ref{fig:fig2}.}
\label{fig:fig3}
\end{figure}

On the other hand, the nonlinear effect of atom-atom interaction can neutralize the effect of CK nonlinearity. In order to illustrate this procedure, in Fig.\ref{fig:fig3} the mean number of intracavity photons has been plotted versus the normalized detuning $ \delta_{c}/\kappa $ for two values of $ \omega_{sw}=5\omega_{R} $ [Fig.\ref{fig:fig3}(a)] and $ \omega_{sw}=10\omega_{R} $ [Fig.\ref{fig:fig3}(b)] when the pump rate has been fixed at $ \eta=2\kappa $. The red (blue) line corresponds to the absence (presence) of the CK nonlinearity. As is seen, for a fixed pump rate by increasing the \textit{s}-wave scattering frequency from $ \omega_{sw}=\omega_{R} $ in Fig.\ref{fig:fig2}(b)  to $ \omega_{sw}=5\omega_{R} $ in Fig.\ref{fig:fig3}(a) the two curves get nearer to each other. Finally, by increasing the \textit{s}-wave scattering frequency up to $ \omega_{sw}=10\omega_{R} $ in Fig.\ref{fig:fig3}(b) the CK effect is completely neutralized by the atom-atom interaction and therefore the two curves overlap completely.

\begin{figure}[ht]
\centering
\includegraphics[width=2.8in]{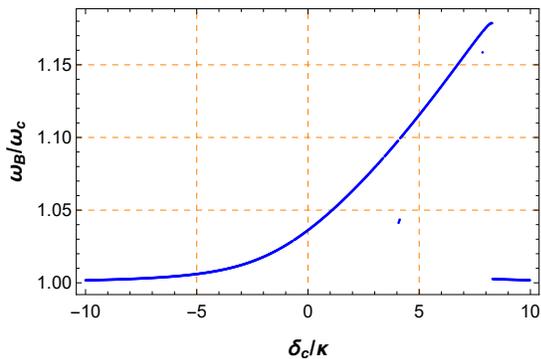} 
\caption{(Color online) The normalized effective Bogoliubov frequency ($ \omega_{B}/\omega_{c} $) versus the normalized detuning $ \delta_{c}/\kappa $ for $ \eta=2\kappa $ and $ \omega_{sw}=\omega_{R} $. The other parameters are the same as those in Fig.\ref{fig:fig2}.}
\label{fig:fig4}
\end{figure}

In the previous section, it was shown how the presence of the CK nonlinearity causes the effective frequency of the  Bogoliubov mode ($ \omega_{B} $) to be dependent on the mean number of interacavity photons [Eq.(\ref{wB})]. Due to the dependence of $ |\alpha|^{2} $ on the effective cavity detuning $ \delta_{c} $ (as depicted in Figs.\ref{fig:fig2} and \ref{fig:fig3}), the effective Bogoliubov frequency also depends on $ \delta_{c} $. In Fig.\ref{fig:fig4} the normalized effective Bogoliubov frequency ($ \omega_{B}/\omega_{c} $) has been plotted versus the normalized detuning $ \delta_{c}/\kappa $ for the external laser pump rate $ \eta=2\kappa $ and $ \omega_{sw}=\omega_{R} $. As is seen from Fig.\ref{fig:fig4}, far away from the bistability region where $ |\delta_{c}|\gg\kappa $ the effective Bogoliubov frequency ($ \omega_{B} $) gets near to the Bogoliubov frequency in the absence of the CK nonlinearity ($ \omega_{c} $) while near the bistability region where $ 3\kappa<\delta_{c}<9\kappa$ the effective Bogoliubov frequency has the maximum deviation from $ \omega_{c} $ because in this region the mean field of the optical mode has the maximum value as has been shown in Fig.\ref{fig:fig2}(b).

\begin{figure}[ht]
\centering
\includegraphics[width=2.8 in]{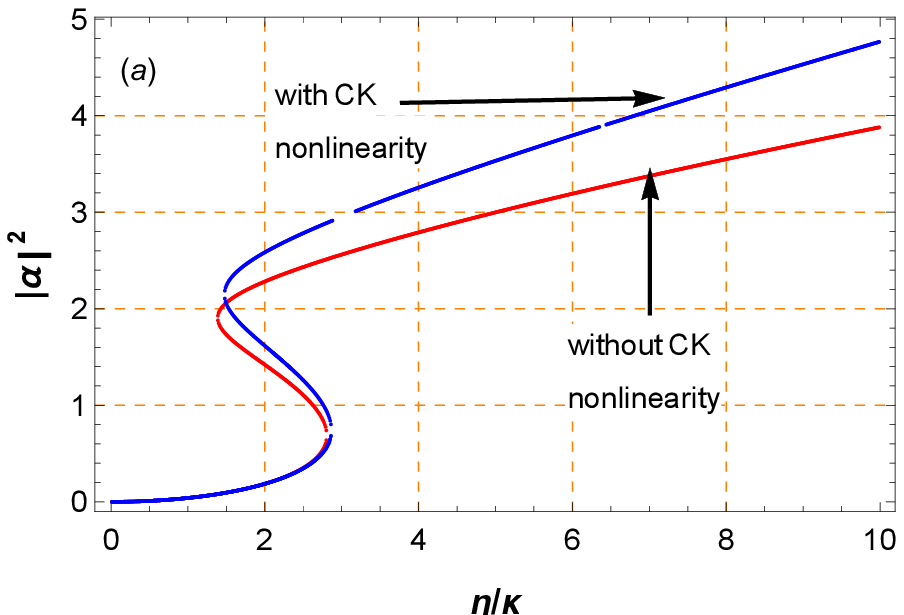}
\includegraphics[width=2.8 in]{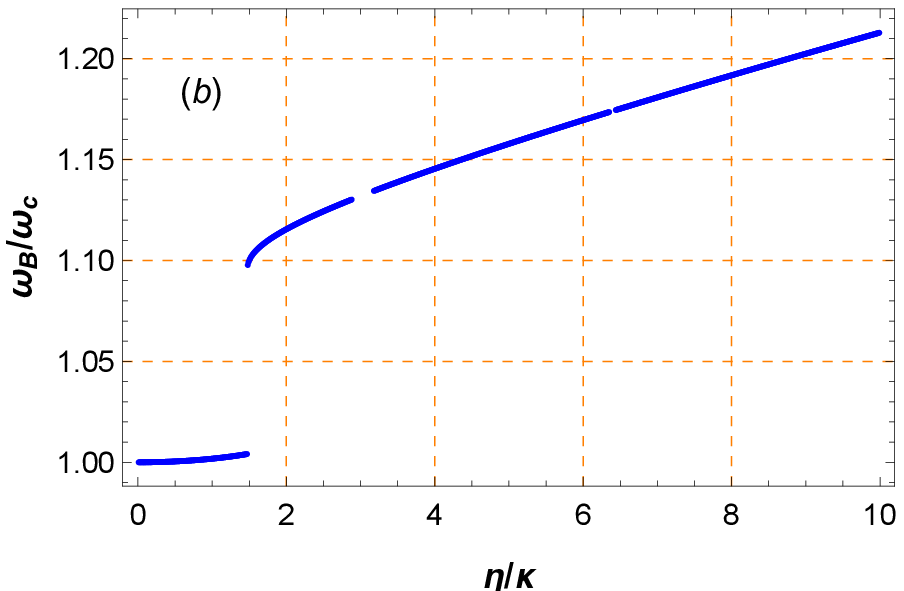}
\caption{
(Color online) (a) The mean number of intracavity photons versus the normalized pump laser rate $ \eta/\kappa $ . The red (blue) line corresponds to the absence (presence) of the CK nonlinearity. (b) The normalized effective frequency of the Bogoliubov quadrature ($ \omega_{B}/\omega_{c} $) versus the normalized pump laser rate $ \eta/\kappa $. Here, $ \delta_{c}=5\kappa, \omega_{sw}=\omega_{R}  $ and the other parameters are the same as those in Fig.\ref{fig:fig2}.}
\label{fig:fig5}
\end{figure}

In order to examine the bistability behavior in terms of the pump laser rate, in Fig.\ref{fig:fig5}(a) we have plotted the mean number of intracavity photons versus the normalized pump laser rate $ \eta/\kappa $ for $ \delta_{c}=5\kappa $ and $ \omega_{sw}=\omega_{R} $. The red (blue) line corresponds to the absence (presence) of the CK nonlinearity. As is seen, for small values of the pump laser rate when the system is in the lower branch the two curves overlap while for large pump rates when the system is in the upper branch the two curves are resolved from each other. It means that for a specified detuning the effect of CK nonlinearity manifests itself when the driving laser becomes more intensive. That is also why in Fig.\ref{fig:fig2}(a) the two curves overlap. 

In Fig.\ref{fig:fig5}(b) the normalized effective Bogoliubov frequency ($ \omega_{B}/\omega_{c} $) has been plotted versus the normalized laser pump rate $ \eta/\kappa $ for detuning $ \delta_{c}=5\kappa $ and $ \omega_{sw}=\omega_{R} $. As is seen, for $ \eta<1.5\kappa $ the CK nonlinearity does not change the effective Bogoliubov frequency so that $ \omega_{B}\approx\omega_{c} $ (corresponding to the lower branch in Fig.\ref{fig:fig5}(a)). At the threshold of the bistability where $ \eta\approx 1.5\kappa $ the effective Bogoliubov frequency makes a sudden jump and its value increases by 10 percent due to the effect of the CK nonlinearity. Beyond the threshold, the deviation of the effective Bogoliubov frequency from $ \omega_{c} $ increases by increasing the pump rate.

\section{effect of the CK nonlinearity on entanglement and squeezing}
In the previous section we studied the effect of the CK nonlinearity on the mean-field values of the system. In this section we will show how the quantum fluctuations and correlations of the system may be affected by the presence of the CK nonlinearity  when the system reaches to the stationary state. Due to the linearized dynamics of the fluctuations and since all noises are Gaussian the steady state is a zero-mean Gaussian state which is fully characterized by the $4\times4$ stationary correlation matrix (CM) $V$, with components $V_{ij}=\langle \delta u_i(\infty)\delta u_j(\infty)+\delta u_j(\infty)\delta u_i(\infty)\rangle/2 $. Using the quantum Langevin equations [Eq.(\ref{nA})], one can show that $ V $ fulfills the  Lyapunov equation \cite{Genes2008}
\begin{equation}\label{lyap}
AV+VA^T=-D,
\end{equation}
where
\begin{equation}\label{D}
 D=\mathrm{Diag}[\kappa,\kappa,\gamma(2n_{c}+1),\gamma(2n_{c}+1)],
\end{equation}
is the diffusion matrix with $ n_{c}=[\exp(\hbar\omega_{B}/k_{B}T-1)]^{-1} $ as the mean number of thermal excitations of the Bogoliubov mode in the presence of the CK nonlinearity (in the absence of CK, $ \omega_{B} $ should be substituted for $ \omega_{c} $). The Lyapunov equation (\ref{lyap}) is linear in $V$ and can straightforwardly be solved. By solving Eq.(\ref{lyap}) we can obtain the matrix $ V $ which gives us the second-order correlations of the fluctuations.

\begin{figure}[ht]
\centering
\includegraphics[width=2.8in]{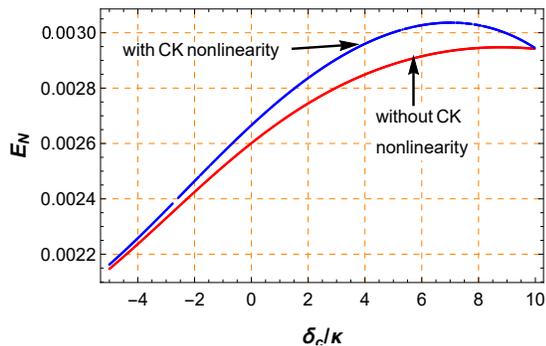} 
\caption{(Color online) Atom-field entanglement versus the normalized detuning $ \delta_{c}/k $ for $ \eta=7\kappa $ and $ \omega_{sw}=\omega_{R} $. The red (blue) line corresponds to the absence (presence) of the CK nonlinearity. The other parameters are the same as those in Fig.\ref{fig:fig2}.}
\label{fig:fig6}
\end{figure}

At first, we examine the atom-field entanglement which is calculated by the logarithmic negativity\cite{eis}:
\begin{equation}\label{en}
E_N=\mathrm{max}[0,-\mathrm{ln} 2 \eta^-],
\end{equation}
where  $\eta^{-}\equiv2^{-1/2}\left[\Sigma(V)-\sqrt{\Sigma(V)^2-4 \mathrm{det}V}\right]^{1/2}$  is the lowest symplectic eigenvalue of the partial transpose of the matrix $V$ which can be written as
\begin{equation}\label{bp}
V=\left(
     \begin{array}{cc}
     \mathcal{A}&\mathcal{C}\\
      \mathcal{C}^{T}&\mathcal{B}\\
       \end{array}
   \right),
\end{equation}
and $\Sigma(V)=\mathrm{det} \mathcal{A}+\mathrm{det} \mathcal{B}-2\mathrm{det} \mathcal{C}$.
 
In Fig.\ref{fig:fig6} the atom-field entanglement has been plotted versus the normalized detuning $ \delta_{c}/\kappa $ in the presence (blue line) and in the absence (red line) of the CK nonlinearity for the pumping rate $ \eta=7\kappa $ and the \textit{s}-wave scattering frequency $ \omega_{sw}=\omega_{R} $, in the the rage of $ \delta_{c} $ where the system is stable. As is seen, the CK nonlinearity increases the atom-field entanglement to some extent for a wide range of the effective detuning $ \delta_{c} $. It is similar to what happens in the bare optomechanical systems where the presence of the CK nonlinearity strengthen the entanglement between the moving mirror and the optical field as has been recently shown in Ref.\cite{Chakraborty}. In the present hybrid system the effect of the CK nonlinearity on the atom-field entanglement is very weak for small pumping rates.

On the other hand, the nonlinear atom-atom interaction, i.e., the fifth term in Eq.(\ref{Hc}), is analogous to the interaction Hamiltonian of a degenerate parametric amplifier (DPA). In a DPA a pump beam generates a signal beam by interacting with a $ \chi^{(2)} $ nonlinearity. This process has long been considered as an important source of the squeezed state of the radiation field \cite{zubairybook}. As has been shown in Ref.\cite{dalafi3}, an interacting BEC behaves as a so-called "atomic parametric amplifier" (APA) in which the condensate acts as an atomic pump field and the Bogoliubov mode plays the role of the signal mode in the DPA. Also, the \textit{s}-wave scattering frequency of atom-atom interaction plays the role of the nonlinear gain parameter. 

\begin{figure}[ht]
\centering
\includegraphics[width=2.8 in]{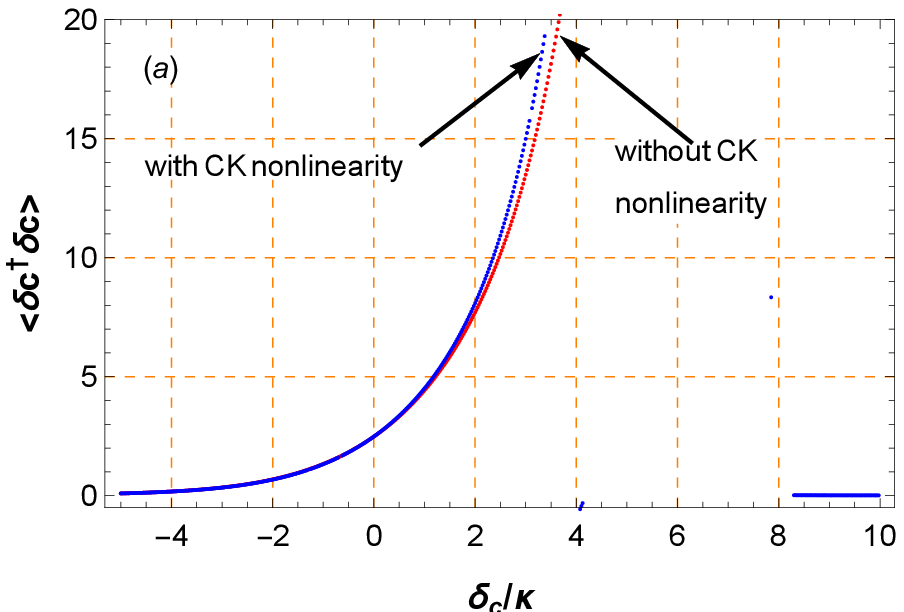}
\includegraphics[width=2.8 in]{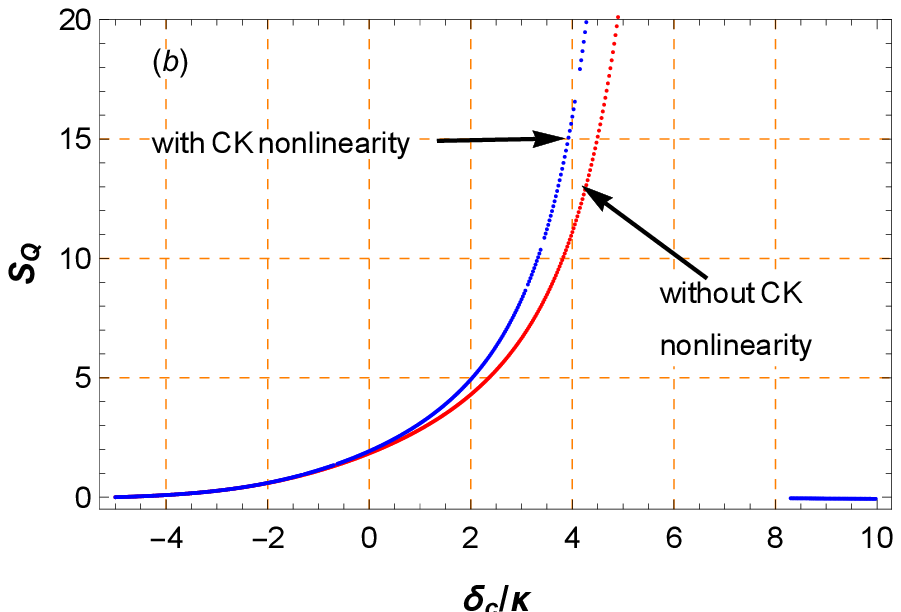}
\caption{
(Color online) (a) the incoherent excitation number of atoms in the Bogoliubov mode and (b) the squeezing parameter of the quadrature $ \delta Q $ versus the normalized detuning $ \delta_{c}/\kappa $. The red (blue) line corresponds to the absence (presence) of the CK nonlinearity. Here, $ \eta=2\kappa, \omega_{sw}=\omega_{R}  $ and the other parameters are the same as those in Fig.\ref{fig:fig2}.}
\label{fig:fig7}
\end{figure}

In the very simplified model studied in Ref.\cite{dalafi3} where the cavity is not driven continuously by a pump laser, it was shown how a squeezed state can be generated disregarding all damping processes.  Here, we would like to show that in a realistic model where the cavity is driven by an external pump laser and in the presence of all damping processes, one can generate a steady-state quadrature squeezing in the Bogoliubov mode of the BEC through the interatomic interactions. In addition, we investigate the effect of the CK nonlinearity on the quantum fluctuations and also on the squeezing behavior of the system.

As is well known, for a single-mode quantum field with quadratures $ q $ and $ p $ obeying  the commutation relation $ [q,p]=i $, the degree of squeezing is defined in terms of the squeezing parameters $ S_{q}=2\langle(\Delta q)^{2}\rangle-1 $ and $ S_{p}=2\langle(\Delta p)^{2}\rangle-1 $ where $ \langle(\Delta q)^{2}\rangle=\langle q^2\rangle-\langle q\rangle^2 $ and $ \langle(\Delta p)^{2}\rangle=\langle p^2\rangle-\langle p\rangle^2 $ are the quantum uncertainties. Whenever $ S_{i}<0  (i=p, q)$, the corresponding state is a squeezed one.

Based on the experimental data of Refs.\cite{Brenn Science,Ritter Appl. Phys. B}, our numerical results show that the stationary squeezing is possible just for the quadrature $ \delta Q $ of the Bogoliubov mode whose squeezing parameter $ S_{Q} $ is calculated from the following equation
\begin{equation}
S_{Q}=2\langle\delta Q^{2}\rangle-1=2V_{33}-1.
\end{equation}
Furthermore, the incoherent excitation number of atoms in the Bogoliubov mode is calulated as
\begin{equation}\label{deltanph}
\left\langle\delta c^{\dagger}\delta c\right\rangle=\frac{1}{2}(V_{33}+V_{44}-1).
\end{equation}

\begin{figure}[ht]
\centering
\includegraphics[width=2.8in]{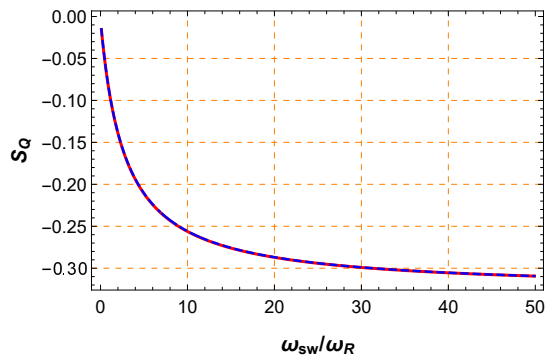} 
\caption{(Color online) The squeezing parameter $ S_{Q} $ of the quadrature $ \delta Q $ of the Bogoliubov mode of the BEC versus the normalized \textit{s}-wave scattering frequency $ \omega_{sw}/\omega_{R} $ for $ \delta_{c}=-15\kappa $ and $ \eta=5\kappa $. The other parameters are the same as those in Fig.\ref{fig:fig2}.}
\label{fig:fig8}
\end{figure}

The incoherent excitation number of atoms in the Bogoliubov mode and the squeezing parameter of the quadrature $ \delta Q $ have been, respectively, plotted in Figs.\ref{fig:fig7}(a) and \ref{fig:fig7}(b) versus the normalized detuning $ \delta_{c}/\kappa $ for $ \eta=2\kappa $ and $ \omega_{sw}=\omega_{R} $ in the presence (blue line) and in the absence (red line) of the CK nonlinearity. As is seen, the quantum fluctuations in the excitation number of the Bogoliubov mode as well as the squeezing parameter $ S_{Q} $ diverge near the bistability region. The presence of the CK nonlinearity increases the slope of the curve and make the fluctuations diverge more rapidly. However, in the dispersive regime where $ |\delta_{c}|\gg\kappa $ the quantum fluctuations fade away so that $ \langle\delta c^{\dagger}\delta c\rangle $ as well as $ S_{Q} $ tend to zero both in the presence and in the absence of the CK nonlinearity.

Therefore, the effects of the CK nonlinearity are appeared in the nondispersive regime where $ |\delta_{c}|\approx\kappa $ while in the dispersive regime neither the mean fields nor the quantum fluctuations are affected by the CK nonlinearity. Instead, the nonlinear atom-atom interaction can change the properties of the system in both regimes.

To verify this result more clarly, in Fig.\ref{fig:fig8}, we have plotted the squeezing parameter $ S_{Q} $ of the quadrature $ \delta Q $ of the Bogoliubov mode of the BEC versus the normalized \textit{s}-wave scattering frequency $ \omega_{sw}/\omega_{R} $ for $ \delta_{c}=-15\kappa $ and $ \eta=5\kappa $. As can be seen, the blue and red lines corresponding to the presence and absence of the CK nonlinearity have a complete overlap because the system is in the dispersive regime. In the range of those values of $ \omega_{sw} $ considered in Fig.\ref{fig:fig8} the system is stable and the Bogoliubov approximation as well as the single mode condition of the BEC is fulfiled. As is seen, the degree of squeezing is increased by increasing the \textit{s}-wave scattering frequency so that for $ \omega_{sw}>30\omega_{R} $ the degree of squeezing reduces to $ S_{Q}<-0.3 $.

Although the nonlinear atom-atom interaction leads to squeezing of the matter field of the BEC, it cannot cause the quadrature squeezing of the optical field of the cavity. However,one can control the squeezing of the Bogoliubov mode of the BEC by the \textit{s}-wave scattering frequency of atom-atom interaction which itself is controllable through the transverse trapping frequency $\omega_{\mathrm{\perp}}$ \cite{Morsch}. Since the quadrature squeezing occurs in the dispersive regime, it is not affected by the CK nonlinearity which manifests itself jut in the nondispersive regime.

\section{Conclusions}
In conclusion, we have studied a driven optical cavity containing a cigar-shaped BEC. In the weakly interacting regime, where just the first two symmetric momentum side modes are excited by the fluctuations resulting from the atom-light interaction, the BEC can be considered as a single mode quantun field in the Bogoliubov approximation. In this way, the Bogoliubov mode of the BEC is coupled to the optical field in two ways: the radiation pressure interaction and the CK coupling.

Although the CK coupling is much weaker than those of the radiation pressure and the atom-atom interactions, it manifests its effect when the intracavity optical field is strong enough (when the driving rate of the cavity is increased). The most important outcome of the CK nonlinearity is that the Bogoliubov frequency of the BEC is dependent on the mean value of the optical field which leads to a change of bistability behavior. On the other hand, the nonlinear effect of atom-atom interaction neutralizes the CK effect when the \textit{s}-wave scattering frequency is increased.

Furthermore, the CK nonlinearity causes the quantum fluctuations of the system to be increased more rapidly near the bistability region. However, in the dispersive regime where the CK effect disappears, a steady-state quadrature squeezing in the Bogoliubov mode of the BEC can be generated by increasing the \textit{s}-wave scattering frequency of the atom-atom interaction.

\section*{Acknowledgement}
A.D wishes to thank the Laser and Plasma Research Institute of Shahid Beheshti University for its support.

\bibliographystyle{apsrev4-1}

\end{document}